\documentclass[apj]{emulateapj}       

\usepackage{graphicx,color}
\usepackage{amssymb,txfonts}
\usepackage{epsfig,natbib}
\usepackage{multirow}
\usepackage{rotating}
\usepackage{sidecap}
\usepackage{hyperref}

\newcommand{\CII}{\ion{C}{2}}
\newcommand{\CIV}{\ion{C}{4}}
\newcommand{\CaII}{\ion{Ca}{2}}
\newcommand{\FeIX}{\ion{Fe}{9}}
\newcommand{\FeXII}{\ion{Fe}{12}}
\newcommand{\FeXVIII}{\ion{Fe}{18}}
\newcommand{\HeII}{\ion{He}{2}}
\newcommand{\SiIV}{\ion{Si}{4}}
\newcommand{\MgII}{\ion{Mg}{2}}

\DeclareMathAlphabet{\mathitbf}{OML}{cmm}{b}{it}
\DeclareMathAlphabet{\mathf}{OML}{cmm}{c}{sl}

\newcommand{\kms}{km~s$^{-1}$}

\slugcomment{Draft \today}
\shorttitle{Transition region/coronal signatures of penumbral jets}
\begin{document}

\title{Transition-Region/Coronal Signatures and Magnetic Setting of Sunspot Penumbral Jets: {\it Hinode} (SOT/FG), Hi-C and {\it SDO}/AIA Observations}

\author{Sanjiv K. Tiwari\altaffilmark{1}, Ronald L. Moore\altaffilmark{1}, Amy R. Winebarger\altaffilmark{1}, Shane E. Alpert\altaffilmark{2}} 
\email{sanjiv.k.tiwari@nasa.gov}
\altaffiltext{1}{NASA Marshall Space Flight Center, Mail Code ZP 13, Huntsville, Alabama 35812, USA}
\altaffiltext{2}{Department of Physics and Astronomy, Rice University, Houston, TX, 77005, USA}
\begin{abstract}
Penumbral microjets (PJs) are transient narrow bright features in the chromosphere of sunspot penumbrae, first characterized by \cite{kats07} using the \CaII\ H-line filter on {\it Hinode}'s Solar Optical Telescope (SOT). It was proposed that the PJs form as a result of reconnection between two magnetic components of penumbra (spines and interspines), and that they could contribute to the transition region (TR) and coronal heating above sunspot penumbrae. We propose a modified picture of formation of PJs based on recent results on internal structure of sunspot penumbral filaments. Using data of a sunspot from {\it Hinode}/SOT, High Resolution Coronal Imager, and different passbands of the Atmospheric Imaging Assembly (AIA) onboard the {\it Solar Dynamics Observatory}, we examine whether PJs have signatures in the TR and corona. We find hardly any discernible signature of normal PJs in any AIA passbands, except a few of them showing up in the 1600 \AA\ images. However, we discovered exceptionally stronger jets with similar lifetimes but bigger sizes (up to 600 km wide) occurring repeatedly in a few locations in the penumbra, where evidence of patches of opposite polarity fields at the tails of some penumbral filaments is seen in Stokes-V images. These large tail PJs do display signatures in the TR. Whether they have any coronal-temperature plasma is ambiguous. We infer that none of the PJs, including the large tail PJs, directly heat the corona in ARs significantly, but any penumbral jet might drive some coronal heating indirectly via generation of Alfv\'en waves and/or braiding of the coronal field. 

\end{abstract}


\section{INTRODUCTION}\label{intro}
\cite{kats07} first reported and characterized penumbral microjets (PJs) from high-spatial-resolution, high-cadence \CaII\ H-line chromospheric images of sunspots obtained by the Solar Optical Telescope/Filtergraph \cite[SOT/FG:][]{ichi08} onboard the {\it Hinode} \citep{kosu07} satellite. The \CaII\ H-line is formed in the chromosphere at a temperature of about 10$^4$ K or less. PJs are scattered, constantly occurring, transient jet events in sunspot penumbrae, with lifetimes of less than a minute, lengths typically between 1000 and 4000 km, some can be longer up to 10000 km, and with widths of less than 600 km. Their apparent speed is more than 100 \kms. Because they are faint and transient, with an enhanced brightness of 10--20\% as compared to background penumbra, PJs are more clearly visible in running difference images in comparison to direct intensity images.

The fact that PJs were more clearly visible in the limbward side of the penumbra and least visible in the disk-center side of the penumbra, evidently due to foreshortening, led \cite{kats07} to conclude that the jets are aligned to the more vertical component of the penumbral magnetic structure. The penumbral magnetic field is a combination of spines (more vertical field) and interspines (more horizontal field), first classified by \cite{lites93}; see also \cite{titl93,sola93,pill00,sola03,bell04,lang05,borr11,scha11,scha12,tiw13,tiw15aa} for further details on the two components of sunspot penumbral magnetic field. \cite{jurc08} observed an increase in the inclination of PJs towards the outer edge of penumbra from an average inclination of PJs of 35$^\circ$ (with respect to the local normal line) at the umbral-penumbral boundary to 70$^\circ$ at the penumbral/quiet Sun boundary.  This inclination change of PJs is compatible with the average field inclination change with radial distance in sunspots, see e.g., Figure 7 of \cite{tiw15aa} for radial variation of field inclination in a sunspot.  

From a space-time plot, \cite{kats07} found some PJs to form near bright penumbral grains \citep{mull73,sobo99,rimm06}. However, this remained to be established statistically. \cite{jurc08} found it difficult to locate PJs with respect to penumbral filaments. \cite{kats07} proposed that PJs could be produced as a result of magnetic reconnection between the two magnetic components (spines and interspines). In this picture, the jets are oriented in the direction of the spine field, travel upward along it, and are rooted at an edge of a filament, between the field in a spine and the more horizontal field (along the central axis) in the filament.

Recent magnetohydrodynamic (MHD) simulations support the idea of magnetic reconnection driving these events, either induced by strong outflows along horizontal flux tubes \citep{saka08}, or by assuming the horizontal field in a twisted flux tube \citep{maga10}. An alternative is given by \cite{ryut08}, in which shocks caused by reconnection between neighboring penumbral filaments can produce PJs in a manner that appears consistent with the observations of \cite{rear13}.       

In a recent investigation of internal structure of a sunspot's penumbral filaments by applying a spatially coupled depth-dependent inversion code \citep{van12,van13} on spectropolarimetric data of Hinode (SOT/SP), \cite{tiw13} found that the penumbral filaments (interspines) resemble elongated convection cells, and behave as stretched magnetized granules. The bright penumbral grains are the heads of penumbral filaments (the head is the filament's end closer to the umbra). Strong upflows, with a field of the polarity of that of the umbra, are observed in the heads of penumbral filaments. The upflow continues along the horizontal axis of the filament for more than half its length outward, and weakens with length. A stronger downflow is observed at the tails of filaments, which contain field of opposite polarity to that of the umbra and heads of filaments. Weak downflows are also observed at the edges of filaments along the length on the sides of the upflows. These lateral downflows also often contain field of opposite polarity to that of the umbra and heads of filaments. This opposite polarity field was found in one third of the total filaments studied by \cite{tiw13}. Also see, \cite{scha13} and \cite{ruiz13} for observational reports on the presence of opposite polarity field in the sides of penumbral filaments, which was also obtained in three-dimensional magnetohydrodynamic (3D MHD) simulations by \cite{remp12}. Thus, keeping the above picture of a penumbral filament's internal structure in mind, the scenario of formation of PJs \cite[proposed by][]{kats07} e.g., by reconnection between two components of penumbral magnetic field, inclined to each other at an acute angle, should be modified. In Section \ref{sec1}, we present a modified picture of the magnetic configuration for the production of PJs.

The other main issue that we address in this paper is whether PJs have any transition region (TR) or coronal signatures. The estimation of chromospheric thermal energy (3/2 $n k_BTV$) of a PJ as estimated by \cite{kats07} based on the following numbers: n = 10$^{18}~ \mbox m^{-3}$, k$_B$ = 1.38$\times 10^{-23} ~\rm{m^2~ kg ~s^{-2}~ K^{-1}}$, T = 10$^4$ K, $V$ = 2000 km $\times$ (300 km)$^2$, returns a value of 2$\times$10$^{16}$ J, or 2$\times$10$^{23}$ erg, which is of the order of that of a coronal nanoflare \citep{sves76,parker88}.  Based on this estimation it was suggested by \cite{kats07} that PJs have potential to appear or display some signatures in the corona, and to contribute in some ways to the coronal heating in active regions. 


To test the above hypothesis, one needs coronal observations of sunspot penumbrae at very high spatial resolution because, as mentioned before, the widths of these PJs are in the range 150 -- 600 km (400 km or less according to \cite{kats07}), close to and at the the resolution limit of the telescope (SOT). Coronal observation at such a high resolution was not available until the High-resolution Coronal Imager \cite[Hi-C:][]{koba14,cirt13} obtained images of an active region (AR) in a narrow passband filter centered at 193 \AA\ at a high spatial resolution of about 150 km, which is also the approximate resolution of Hinode (SOT/FG). Fortunately, as a result of coordinated observations, Hi-C and Hinode both observed a part of a penumbra of the Hi-C AR 11520 for 1.75 minutes. Although the part of the observed penumbra was on the disk-center side, which is not the best location for the visibility of PJs, the penumbral field was twisted far from the disk-center direction, so that we could observe several PJs in this part of penumbra within this short time period of observation. 

We have extended the study for one hour by using different Atmospheric Imaging Assembly channels. We notice a few locations where larger jets are repeatedly produced. To investigate the magnetic structure and origin of these larger penumbral jets, we have also used the co-temporal Stokes-V data obtained by SOT/FG in the field of view (FOV) of the SOT/FG \CaII\ movie and G-band images of the penumbra.

\section{DATA}\label{data}

To identify PJs in the penumbra of the NOAA AR 11520 ($\sim$ X $-$125\arcsec, Y $-$325\arcsec) observed by Hi-C \citep{cirt13,koba14}, we have used \CaII\ H-line broadband filtergraph  images (centered at 3968 \AA) obtained by SOT \citep{tsun08,suem08,ichi08} onboard {\it Hinode} \citep{kosu07}. We use observations taken between 18:53 and 19:53 UT on 11 July 2012, which covers 1.75 minutes (18:53:44 -- 18:55:30 UT) overlap with the five minutes observations of Hi-C. Hi-C observed at high-resolution the corona of AR 11520 in a narrow wavelength range centered at 193 \AA. The cadence of the Hi-C 193 \AA\ images is 5 s. The cadence of the \CaII\ images varies frame to frame from 5 s to 15 s. The spatial resolution of both telescopes is about 150 km. 

G-band images from SOT/FG are used to identify the locations of feet of PJs in the photosphere. Heads of filaments appear brighter and tails darker in G-band images. We have also used Stokes-V images obtained by SOT/FG to examine the magnetic polarity at the locations of some larger jets, discussed later. Both the G-band and Stokes-V images have a cadence of about 50 s.

\begin{figure*}
      \centering
      \includegraphics[width=\textwidth]{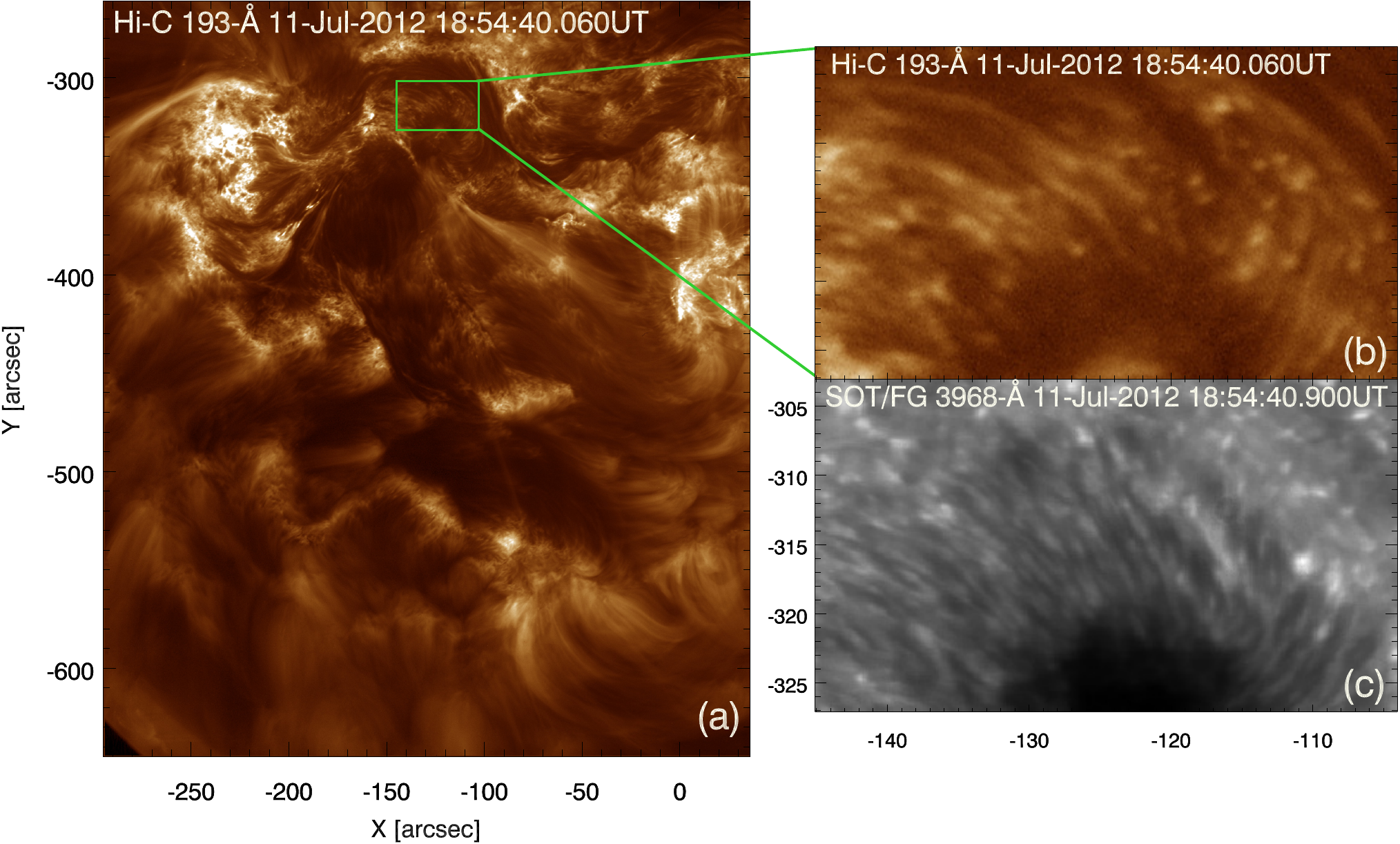}
 \caption{Overview of observations by Hi-C and Hinode (SOT/FG): (a) Hi-C full field of view 193 \AA\ image; a box covers the region of interest of sunspot penumbra for the present work. (b) Enlargement of area outlined by the box in (a). (c) Hinode (SOT/FG) Ca {\footnotesize II} H-line image of the same FOV at the same time as the Hi-C image. A movie `movie1.mp4' of 1.75 minutes is linked to this figure. The movie contains four panels: two top panels are of Hi-C 193 \AA\ filtergrams and SOT/FG \CaII\ H-line filtergrams, and two lower panels are of running differences of the corresponding intensity images. Solar north is up and west is to the right in these and all other solar images/movies in this paper.}
      \label{f2} 
\end{figure*}

In addition, we use one-hour-long (18:53 -- 19:53 UT) image series from 1600, 304, 171, 193, and 94 \AA\ passbands observed by the Atmospheric Imaging Assembly \citep[AIA:][]{leme12} onboard the {\it Solar Dynamics Observatory (SDO)} spacecraft. The pixel size of AIA is 0.6 arcsec, so these images detect but do not resolve the jets. The cadence of the AIA 1600 channel is 24 s, and the cadence is 12 s for each of the other AIA channels used in the present study. All the data used in the paper are calibrated and, whenever applicable, co-registered by using SolarSoft routines.

The 193 \AA\ (AIA and Hi-C) and AIA 94 \AA\ bands are both predominantly coronal \citep{leme12}. The 193 \AA\ band particularly detects \FeXII\ at about 1.5 MK, but also has some response to the transition region emission of $2-3 \times10^5$ K plasma \citep{delz13,wine13}. The 94 \AA\ channel allows mostly hot emissions and is centered on an \FeXVIII\ line ($6-8\times10^6$ K), but also detects some line emission from Fe ions formed at $\sim 1\times 10^6$ K \citep{warr12,delz13}; see also \cite{test12}. There is no known cool TR contamination in the 94 \AA\ channel.  

The 1600 \AA\ AIA passband, which primarily passes lower-chromospheric continuum emission, also covers the two \CIV\ lines near 1550 \AA\ formed at T $\approx$ 10$^5$ K in the transition region (TR). The transient brightenings in the 1600 \AA\ band are due to these \CIV\ lines, and hence are from the lower TR \citep{leme12}. The 304 \AA\ passband  observes the upper chromosphere and lower TR at $\sim$ 50,000 K, in emission primarily from \HeII. The 171 \AA\ band observes emission primarily from \FeIX\ formed in the upper TR at 6 $\times 10^{5}$ K.

\section{RESULTS}
In this section, we first present results on detection of TR signatures but no definite coronal signatures of PJs from the analysis of Hinode (SOT/FG) and Hi-C data sets. Then we present results from analysis of Hinode (SOT/FG) and SDO/AIA data of one hour. We discover larger penumbral jets than normal ones reported earlier by \cite{kats07}. We characterize these larger PJs and their magnetic setting.

\begin{figure*}[htp]
      \centering
      \includegraphics[width=\textwidth]{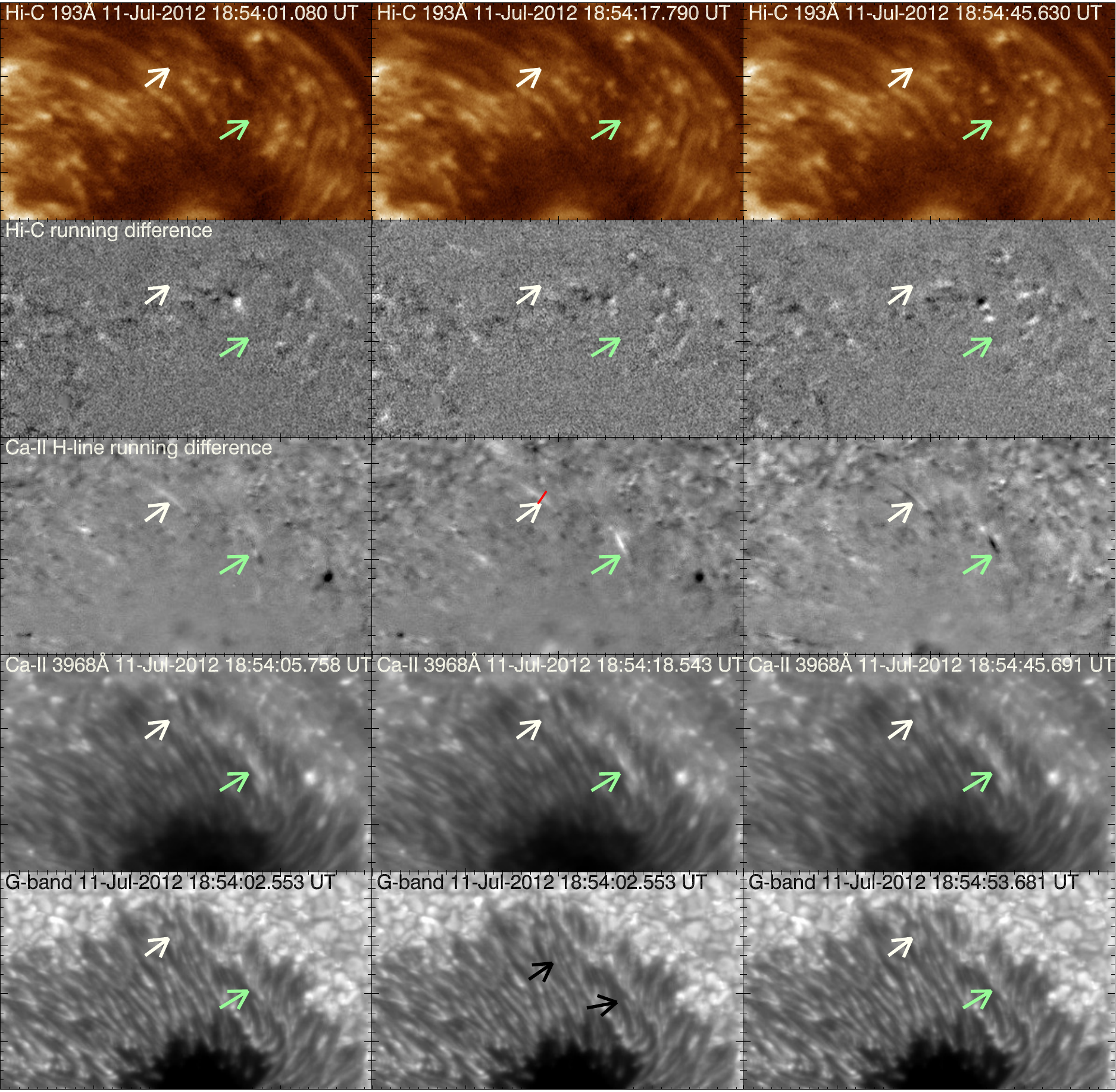}
 \caption{Two examples of penumbral jets, from their birth to decay, are displayed in the middle row. White arrow points to a PJ whereas pale green arrow points to a location where larger PJs are produced repeatedly, discussed in section \ref{trc2}. Top two rows contain Hi-C images and corresponding running difference (RD) images. Third and fourth rows contain RD and intensity images from \CaII\ H-line of nearly the same time as that of the Hi-C image. Last row contains G-band images corresponding to beginning and end of the two jets; the first two G-band images are the same and duplicated to point out the penumbral filament head at the feet of the PJ indicated by white arrow. These arrows are duplicated in all images, except in the middle G-band image, which contains two black arrows, pointing to the source locations (feet) of the jets. A red cut on the middle \CaII\ RD image is to measure the PJ's width, presented in Figure \ref{pjw}. The tickmark separation in each panel is 1\arcsec.} 
\label{f3} 
\end{figure*}

\subsection{Transition region and coronal signatures of penumbral jets: Hinode (SOT/FG) and Hi-C observations}\label{trc1}

In Figure \ref{f2}(a), we display a full FOV Hi-C 193 \AA\ image with a box on it outlining the FOV of the penumbra studied, which is enlarged in panel (b). A co-temporal \CaII\ image of the same FOV is shown in panel (c). A movie `movie1.mp4', connected to this figure, contains four panels with two intensity panels of Hi-C and \CaII, respectively, and two panels of corresponding running difference images, for 1.75 minutes (during 18:53:44 -- 18:55:30) when Hi-C and FG simultaneously looked at the sunspot penumbra. As mentioned before, both instruments have a spatial resolution of about 150 km, which provides a unique opportunity for comparison of the jets at two different heights and temperatures.

At least 10 jets can be observed in the movie `movie1.mp4' of 1.75 minutes; they are especially visible in the \CaII\ running difference movie. Most of the PJs have brightness enhancement by 10 -- 20\% of penumbral background, however some of the largest jets have enhancement up to 30\%. In the Hi-C 193 \AA\ movie and the corresponding running difference movie, one can notice many bright dots (BDs) that move toward or away the spot center along filaments in the penumbra. These BDs are reported and characterized in detail by \cite{alpert15}. However no signatures of PJs are noticeable in the Hi-C 193 \AA\ movie.

In Figure \ref{f3} we display two examples of jets from start to end, clearly seen in the running difference \CaII\ images (middle row). To look for any coronal signatures of them, corresponding Hi-C images and their running difference images are shown in the two upper rows. We have also displayed \CaII\ intensity and G-band images in the last two rows of the Figure \ref{f3}. As noted from the arrows in G-band images it is difficult to identify exactly where PJs are rooted relative to heads of penumbral filaments; this agrees with \cite{jurc08}. 
Only two G-band images, corresponding to beginning and decay times of the two PJs, are available. The middle G-band image is the same as in the first column and black arrows on it point to the locations in the penumbra where the example PJs are produced. Although many of the PJs appear forming near heads of filaments, as pointed by a black arrow in the middle-panel G-band image in Figure \ref{f3}, as an example, it cannot be concluded with certainty that most PJs form at heads of filaments. The second jet pointed by the pale green arrow is among one of the larger penumbral jets, which appear to be rooted at the converging tails of a few filaments (location pointed by a black arrow in the middle G-band image), discussed in detail in section \ref{trc2}.  

Both chromospheric penumbral jets displayed in Figure \ref{f3} show no signatures in the Hi-C running difference images. The same is true for all other PJs seen in the movie `movie1.mp4'.  

In the RD movie of FG \CaII\ images, we notice a couple of locations where somewhat larger jets repeatedly appear. One such location is pointed by the pale green arrow in Figure \ref{f3}. But less than two minutes is too short a time to confirm this repetition. To explore in detail if jets are repeatedly produced at this location for longer time, and if there are other such locations of penumbral larger jets, and whether they have any TR or coronal signatures, we extended our investigation for about an hour of SOT's \CaII\ H-line observations of the penumbra. We included SDO/AIA data to look for TR and coronal signatures of larger PJs, as described next (in Section \ref{trc2}).      

\subsection{Transition region and coronal signatures of larger penumbral jets:  Hinode (SOT/FG) and SDO/AIA observations}\label{trc2}

In the one-hour \CaII\ H-line running difference (RD) movie, we noticed at least four locations where multiple jets are produced repeatedly. These jets appear brighter, larger and wider than most PJs. In what follows, we characterize these larger penumbral jets and investigate their signatures in the TR and corona. We also look for the formation mechanism of these larger PJs by studying magnetic field polarity using Stokes-V images, which is described in Section \ref{mag}. 

\begin{figure*}[htp]
      \centering
      \includegraphics[width=0.91\textwidth]{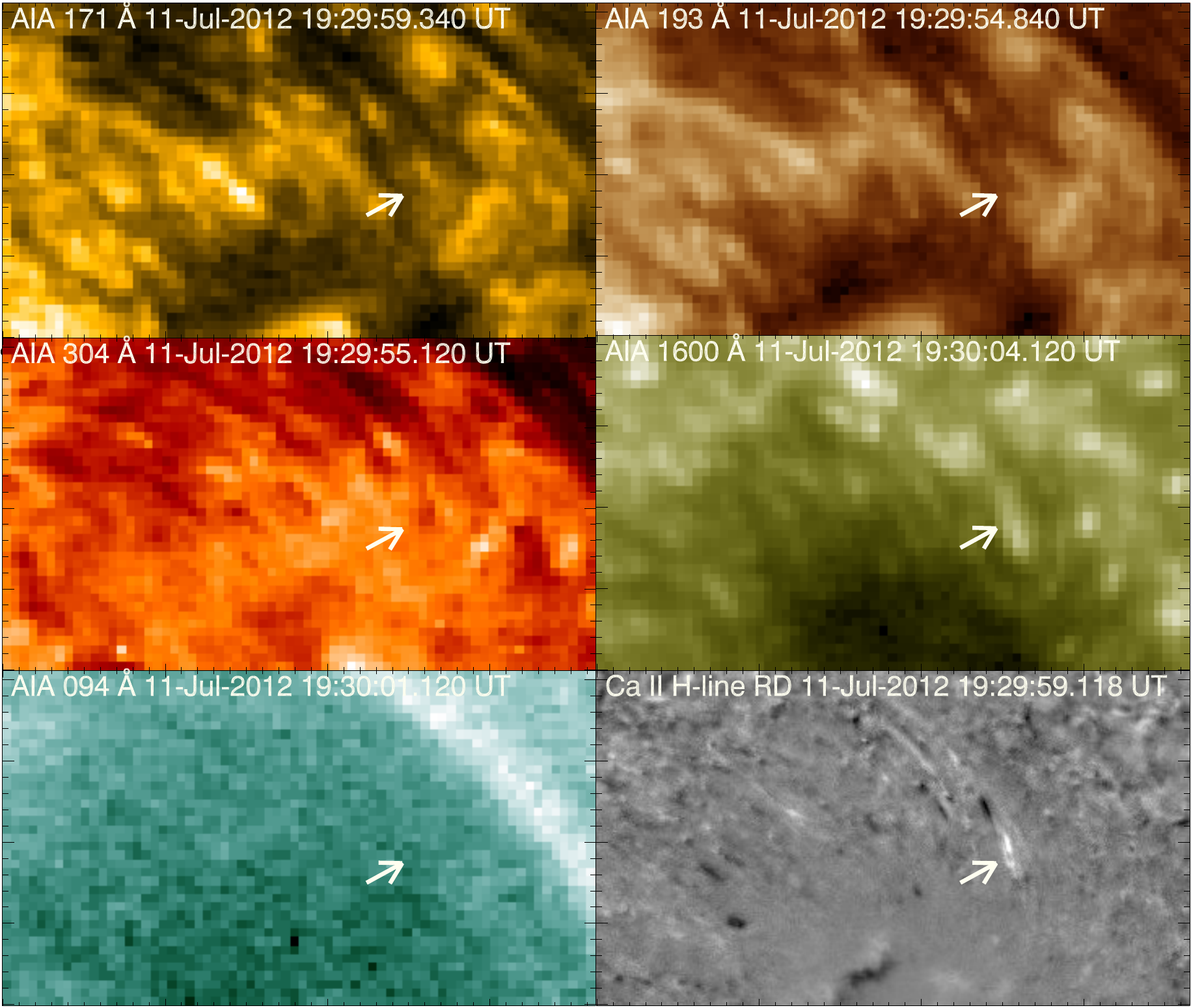}
 \caption{An example image of each wavelength of AIA used in this investigation. The nominal wavelength of the AIA channel, and time of observation, which is nearly the same for all, is displayed on each image. Bottom right panel is a SOT/FG running difference (RD) image, taken closest to the time of all AIA images. To better visualize any transient events, i.e., jets, we have linked a movie `movie2.mp4' of RD images for one hour. White arrows indicate the position of a large PJ (location numbered `4' in the movie). The tickmark separation in each panel is 1\arcsec.} 
      \label{f4} 
\end{figure*}

\begin{figure*}[]
      \centering
      \includegraphics[width=0.91\textwidth]{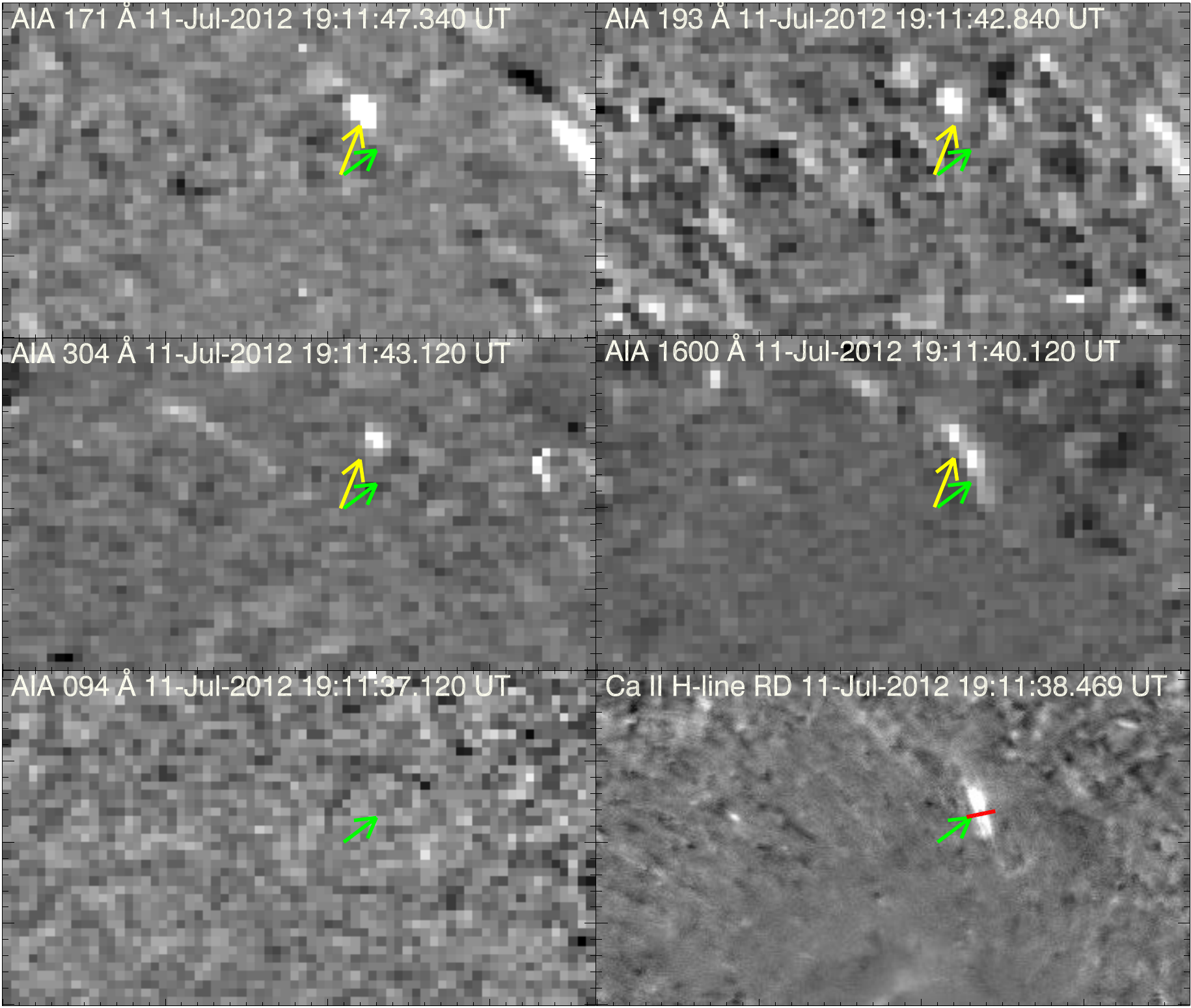}
 \caption{An example of a large PJ (location numbered `2' in the `movie2.mp4'). RD images of five AIA channels and \CaII\ H-line at nearly the same time are shown. The green arrow in each panel marks the location of the example jet as seen in the \CaII\ H-line RD image. The yellow arrow points to the extension of the jet seen in AIA 193, 304 and 171 \AA. A red cut in the bottom right panel is to measure the width of the jet, as presented in the right panel of Figure \ref{pjw}. The tickmark separation corresponds to 1\arcsec.} 
      \label{f5} 
\end{figure*}

First, the AIA image of each wavelength (1600, 304, 171, 193, and 94 \AA) closest in time to each FG \CaII\ H-line image is selected. Then we make both intensity and RD movies of each of these sets of images. In Figure \ref{f4}, we display one example image of all AIA channels used to investigate the signature of penumbral jets in coronal and TR emission. A movie `movie2.mp4' containing six corresponding panels of RD images, in which the jets are better visible, is connected to this figure. The FG \CaII\ RD movie of one hour shows many normal PJs and at least four locations of larger PJs. These four locations of larger PJs are each indicated by an arrow and are numbered in the movie. One of these four locations (numbered `2') is the same as that pointed by pale green arrow in Figure \ref{f3}. The larger PJs have enhanced brightness of 30--60\% to that of the background penumbra.

A clear signature of the example larger PJ is visible in 1600 \AA\ in Figure \ref{f5}. The jet appears extending towards north, higher along the magnetic field, and only the front brightens in 193, 171 and 304 \AA\ channels. No intensity enhancement is noticeable in 94 \AA. Many similar larger jets are observed showing signatures in the four above mentioned wavelengths but no signatures of any of these larger jets were observed in 94 \AA, thus indicating that there is no appreciable MK plasma produced in these penumbral jets, howsoever large they are, and that the brightenings in the other channels is from the emission from transition-region-temperature jet plasma \citep[see, e.g.,][]{wine13}.   

Also noticeable in the movie are some normal PJs, which are neither repetitive nor exceptionally large, with enhanced intensity of 25--30\% to background, displaying faint signatures in 1600 \AA. However signature of these PJs in 304, 171, 193 and of course in 94 \AA\ is hard to detect if at all. This indicates that some larger-end normal PJs do appear in TR emission, but not in coronal emission, in agreement with the absence of their detection in the Hi-C 193 \AA\ movie of 1.75 minutes described in Section \ref{trc1}.   

\begin{figure}[htp]
      \centering
      \includegraphics[width=\columnwidth]{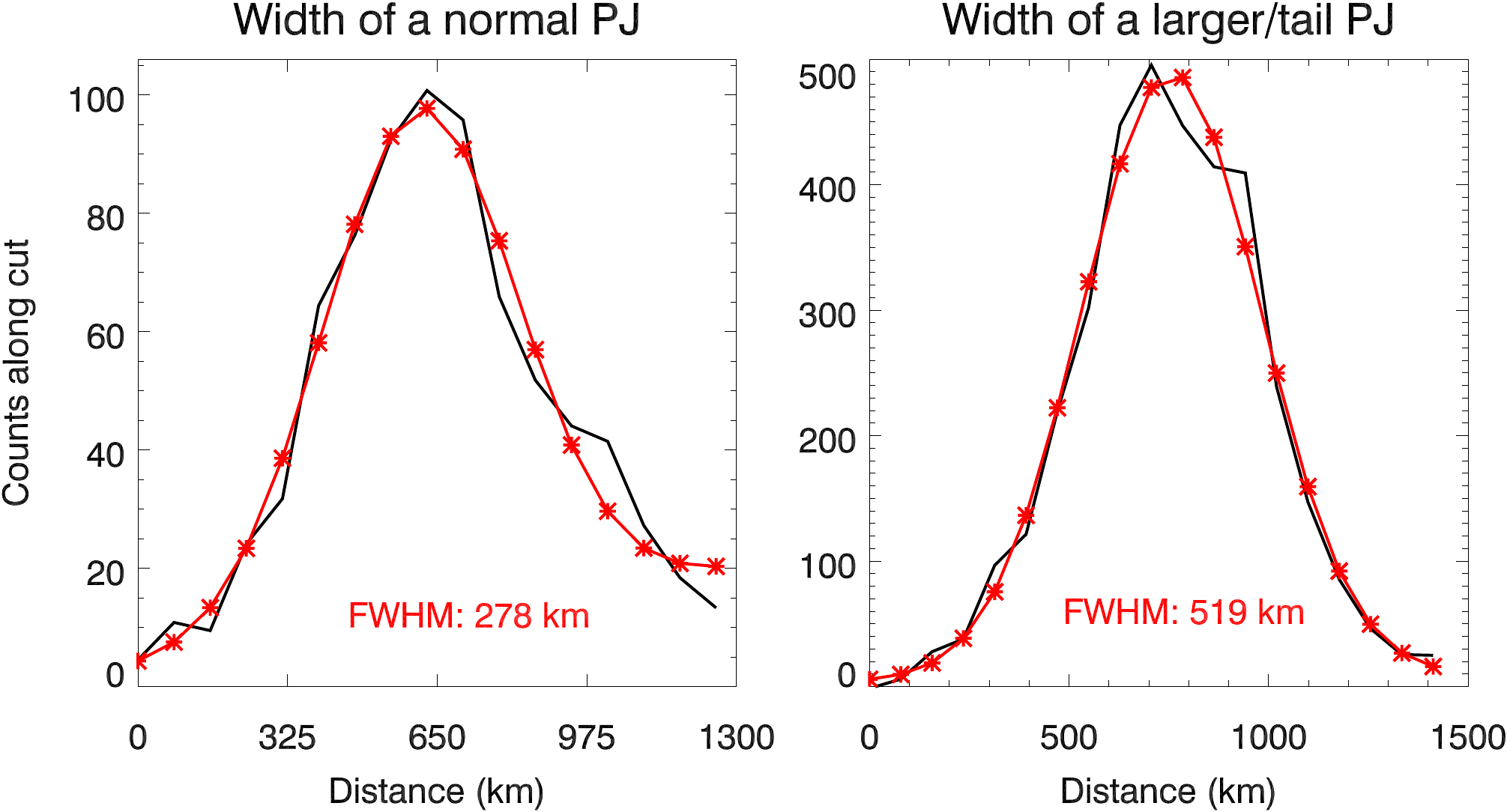}
 \caption{Widths of a normal jet and a larger jet: left panel is for a normal PJ (along red cut in Figure \ref{f3}), right panel is for a larger PJ (along red cut in Figure \ref{f5}). Black and red colored plots show the original intensity and the fitted Gaussian function, respectively. }
      \label{pjw} 
\end{figure}

We measured widths of many of the normal and larger penumbral jets. An example of width of each of them is shown in Figure \ref{pjw}. These examples are ones among the largest and widest in both the cases. Larger PJs can be much wider than normal PJs, as about double in the presented example. Remember the widest normal PJs seen by \cite{kats07} was less than 400 km. The larger PJ shown here is about 500 km wide. When measured by using the method of fitting a Gaussian and taking its full width at half maximum (FWHM), as in Figure \ref{pjw}, the widths of our largest PJ is found to be 600 km. 

The lifetime of larger PJs is however less than a minute, similar to that of normal PJs or a little longer. Larger PJs are repeatedly produced at the same locations (pointed by arrows in the movie `movie2.mp4') in the sunspot penumbra. The length of the largest PJ is 4200 km, however please note that the location of the sunspot (close to disk center), and particularly of its penumbra (disk-center side) does not allow a measurement of the actual length owing to projection. Nonetheless, the estimated lengths of our larger PJs fall in the range of that of the longest PJs (up to 10000 km) observed earlier by \cite{kats07}, indicating larger PJs are of the same category as normal PJs. 

A rough estimation of the speed of the larger penumbral jet at location `4' in `movie2.mov' between the time 19:40:25 and 19:41:03 gives a value of 250 km s$^{-1}$. This speed is much faster than the acoustic speed of 10 \kms\ in the chromosphere. This speed is also larger than that of normal PJs (100-150 \kms) by a factor of two. The Alfv\'en speed in the penumbral chromosphere is of order 1000 \kms. So the speed of the larger PJs is supersonic but sub-Alfv\'enic. Because the deduced observed projected speed is less than the true speed, the fastest larger PJs might be nearly Alfv\'enic.
 
The width of penumbral filaments is about 550 km \citep{tiw13}. Normal PJs appear rooted at an edge of a filament head, and are usually 150-300 km wide; the lower limit of their width is not well known owing to the resolution limit of the SOT. The size of larger PJs, particularly their widths, seems too large to be produced at the same locations as most PJs are. Although it is difficult to detect opposite polarity signal in absence of SP scans, and advanced processing, e.g., done by \cite{tiw13}, we plausibly can detect the signals in Stokes-V images from SOT/FG, especially if larger PJs are produced at a different location having more reversed-polarity flux than the normal PJs. For this purpose, we analyzed corresponding Stokes-V images of one hour, as presented in the next section.

\subsection{Magnetic setting of larger penumbral jets}\label{mag}

\begin{figure*}[t]
      \centering
      \includegraphics[width=\textwidth]{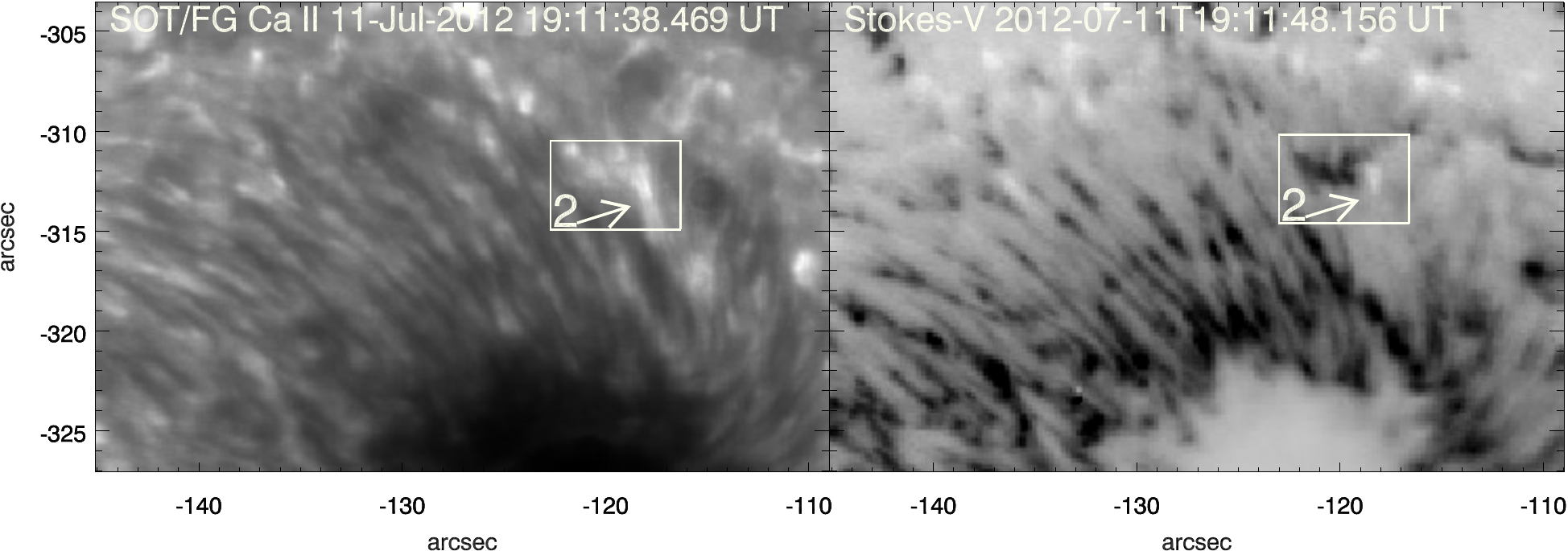}
 \caption{(Left) Same FOV of a \CaII\ H-line image as that of observed Stokes-V images. (Right) A Stokes-V image of nearly the same time as the \CaII\ H-line image. The box in each panel covers a location (numbered `2' in the `movie3.mp4') where larger/tail PJs are produced repeatedly. Absolute flux calculated in arbitrary units, by integrating the Stokes-V signal in the area of the box, is shown in Figure \ref{flux}. }
      \label{stv_ca} 
\end{figure*}

To investigate the magnetic origin of larger PJs, we looked at the Stokes-V data of the same time span as that of the \CaII\ H-line analyzed in this work. The Stokes-V images can be considered equivalent to line-of-sight (LOS) magnetograms, albeit with arbitrary units. The FOV of Stokes-V images is somewhat smaller than that of the \CaII\ H-line images in east-west extent; it covers all but a western strip of the \CaII\ FOV. The cadence of the Stokes-V images is about 50 s, which is poorer than that of the \CaII\ images. In Figure \ref{stv_ca}, we display two images, one of Stokes-V on the right and the other a \CaII\ H-line image of nearly the same time and FOV, on the left. A movie `movie3.mp4' of Stokes-V images is linked to the Figure \ref{stv_ca}. Because of smaller FOV only three of the larger PJ locations can be viewed in this movie. Each of the three locations available in this FOV is marked by an arrow. 

From Figure \ref{stv_ca}, and `movie3.mp4', it is clear that there is presence of opposite polarity magnetic field patches at the locations of these PJs in the Stokes-V images of FG. The opposite polarity patches present in all the three locations of larger PJs's have different scales, forms and visibility. However, from the G-band images and from the Stokes-V movie, each of them appear to be located at or around the tail of a penumbral filament or at a location where tails of several penumbral filaments converge. Tails of filaments have opposite polarity field to that of spines \citep{tiw13}. Therefore the opposite polarity fields in spine and tails can reconnect and produce larger PJs, in a repetitive manner due to presence of the large amount of opposite polarity flux at tails. Occasionally tails of penumbral filaments converge and form stronger patches of field of opposite polarity to that of spines \citep{van13,tiw13}, thus increasing the possibility of both the size and the frequency of larger PJs. 

\begin{figure}[htp]
      \centering
      \includegraphics[width=0.8\columnwidth]{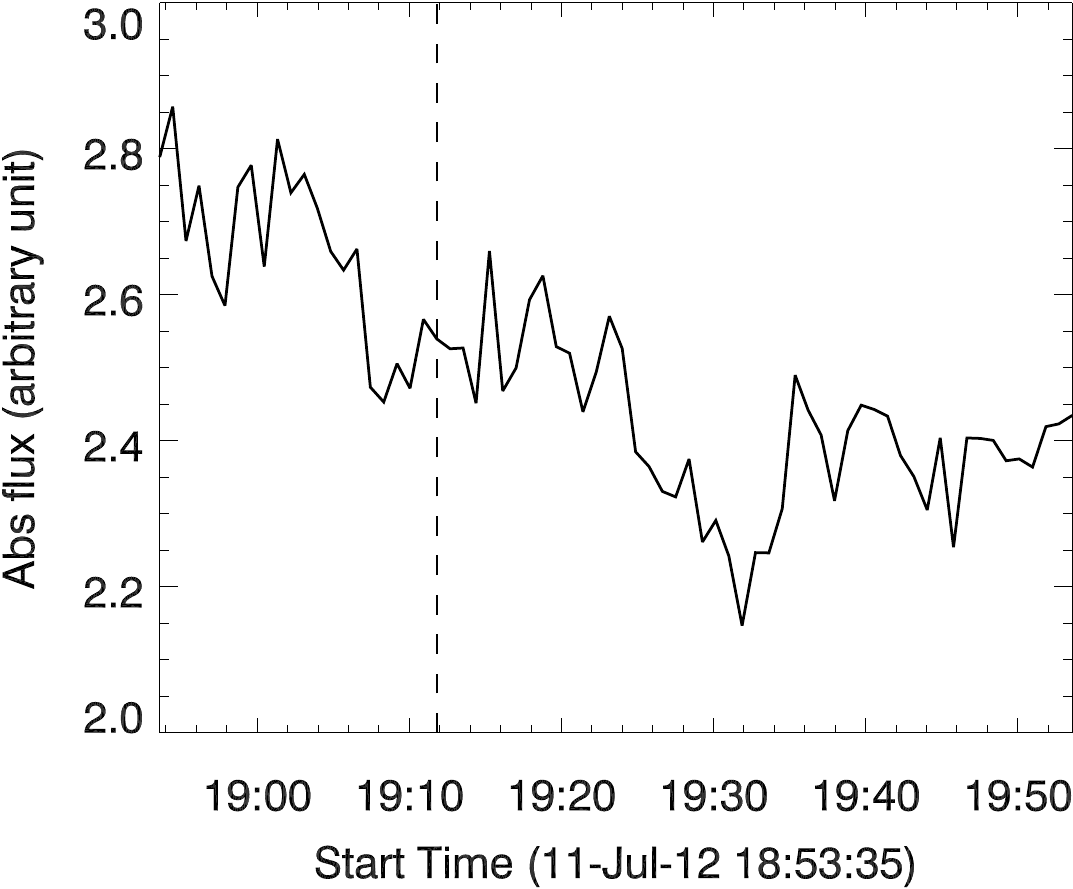}
 \caption{Time plot of absolute flux, inferred from Stokes-V signal, integrated within the box shown in Figure \ref{stv_ca}. Please note that the unit of the absolute flux is arbitrary. A vertical dashed line marks the start time of the example larger PJ indicated by arrow in Figure \ref{f5}. Larger PJs are rooted at this location in the box in Figure \ref{stv_ca}. }
      \label{flux} 
\end{figure}

The locations of larger PJs on the photosphere may also contain emerging flux, cancellation of emerging flux or both. We show one example in Figure \ref{stv_ca} by outlining the area around a location of larger PJs (numbered as `2' in the `movie2.mp4') by a box. We selected the area within the box, and computed absolute Stokes-V signal contained therein. This is equivalent to absolute flux, however, in an arbitrary unit. Figure \ref{flux}, shows the plot of absolute Stokes-V signal. Clearly, a trend of decreasing flux is seen for about first 40 minutes (during which the location `2' produces larger jets) that indicates a cancellation of Stokes-V signal of opposite polarity field. From the movie `movie3.mp4', we notice that the existence of patches of positive polarity, which seem to be the locations of convergence of several filaments, emerge and cancel with the existing negative flux, as larger PJs get produced there. 

Compared to the site `2', similar or smaller patches of opposite polarity fields are observed in the other two locations of larger PJs. These magnetic field patches that have opposite polarity to that of the sunspot umbra appear to be tails of penumbral filaments surrounded by the opposite polarity field of spines or heads of neighboring filaments or both. The bipolar patches could instead be produced by sea-serpent field lines \citep{sain08} that move slowly outward. In either case the reconnection can take place between the field submerging into the photosphere (tails of filaments) and the existing opposite-polarity spine field and/or field inside heads of neighboring filaments. Please remember that the opposite polarity field at the tails of filaments of inner penumbra may not be so clearly visible owing to the Stokes-V signal being dominated by the surrounding spines there.  

These observations suggest that the larger PJs are produced by magnetic reconnection between impacted opposite polarity fields. The polarity arrangement might or might not be compatible with standard jet models, see, e.g., \cite{shib95,anti99,moor01,moor10}.

\section{DISCUSSION}
We have addressed two issues concerning penumbral jets in this paper. First, we have investigated whether PJs have any TR or coronal signatures. During the course of this part of the investigation, we identified locations in the penumbra, apparently near tails of filaments, at which repeatedly occurred PJs larger than the normal PJs. We investigated the magnetic setting of these larger PJs using Stokes-V images and also studied their signatures in the TR and corona. Second, based on the recent observational results on the internal structure of penumbral filaments \citep{tiw13}, and the observations of larger PJs repeatedly appearing close to tails of penumbral filaments in the present paper, we propose a modified picture of the formation of PJs, detailed in Section \ref{sec1}.

\subsection{Transition-region/coronal signatures of penumbral Jets}

Taking benefit of high resolution ($\sim$150 km) of both the instruments the Hinode (SOT/FG) and Hi-C, we compared the \CaII\ H-line and 193 \AA\ images, searching for any signature of chromospheric PJs in the 193 \AA\ coronal images. A comparison of running difference images of 1.75 minutes of both Hinode (SOT/FG) \CaII\ H-line and Hi-C 193 \AA\ reveals no significant signature of the \CaII\ jets in 193 images. However, 1.75 minute was too short period for a firm conclusion. An extension of our investigation by combining SDO/AIA multiple channels, albeit with lower resolutions, reveals some repeating larger jets at some locations. In the G-band images and Stokes-V movie of SOT/FG, these locations appear to be tail of a penumbral filament or where tails of a few filaments converge.

We also detected some normal (non-tail, non-repetitive) PJs clearly displaying signatures in 1600 \AA, but not in 193 \AA\ and other AIA channels studied in this paper. This is in agreement with the recent observations of \cite{viss15}, who found TR signatures of PJs in \CII, \SiIV\ and \MgII k\ slit-jaw images of IRIS data. Please note that the sensitivity and resolution of AIA 1600 \AA\ might have prevented the detection of response of PJs in the TR. Or, the PJs selected by \cite{viss15} all being large enough, in the range of ones the largest normal PJs in our case showing the TR response in 1600, cannot be ruled out. Thus, we conclude that although most of the normal PJs do not directly contribute to the TR and coronal heating, some of the largest normal PJs do display signatures in the TR. If smaller normal PJs do contribute to significant coronal heating in some form, it  could be only by adding energy through increasing stress and braids in coronal loops, or by sending MHD waves up into the corona. 

We observed at least four locations in the FOV of the sunspot penumbra we studied, where larger jets were produced repeatedly. These jets are brighter and larger in size than normal PJs. We find that although the lifetime and length of larger PJs are similar to or at the larger end of the distribution of normal PJs, they are wider, and have higher apparent speeds than the normal PJs. The width of the weakest jets hits the resolution limit of the telescope, SOT, whereas the largest PJs are as wide as 600 km, measured by a Gaussian fit as shown in Figure \ref{pjw}, for an example. The opposite polarity patches found at the base of the larger PJs, which apparently are the locations of tails of filaments, sometimes where multiple filaments tails converge \citep{van13}, suggests a magnetic reconnection process to be responsible for their formation too. Because at tails of filaments there exists larger patches of opposite polarity field than the other parts (head and bulk), a larger scale magnetic reconnection of the field at tails with spine field is plausible. The flux cancellation as shown in Figure \ref{flux} for the example location of larger PJs is compatible with the reconnection being responsible for their production. 

The speeds of all jets, either normal PJs or larger/tail PJs, are faster than sonic but lower than the Alfv\'en speed, consistent with these jets being magnetically driven via reconnection. Thus, both the normal and larger/tail penumbral jets appear to be produced by magnetic reconnection process, in some ways similar to the ways other jets, flares and CMEs are produced, see, e.g., \citet{shib95,anti99,moor01,moor10}, and references therein. 

Although the velocity of larger/tail PJs and their length are on the large side of that of PJs, we are unable to measure their true values due to projection foreshortening in our penumbra. A future investigation with such high resolution data sets, together with Stokes-V images or magnetograms is required to better establish the characteristics of tail PJs.  

Please note that the locations of larger PJs often also produce smaller (normal) size PJs. In other words, not all jets formed at the locations of larger PJs are large enough to appear in the hotter AIA channels (e.g., in 193 \AA). For example, the jet marked by green arrow in the \CaII\ H-line RD images of Figure \ref{f3} is at a location of larger PJs (numbered `2' in the `movie2.mp4') but does not show any signature in 193 \AA\ RD images of Hi-C and/or AIA. Whereas, the jet displayed in Figure \ref{f5} at the same location but at a different time clearly displays signatures in AIA 1600, 304, 171 and 193 channels.   

The estimated thermal energy of chromospheric normal PJs is of the order of 10$^{23}$ erg, and most PJs hardly show any TR or coronal signatures. Whereas, some penumbral jets with larger sizes, that are produced at tails of penumbral filaments, have width of 2-3 times more than that of most PJs, and thus have energy larger by a factor of four to nine, or more, show signatures in the AIA 1600, 304, 171, and 193 \AA\ channels. Because AIA 304, 1600, 171 are formed in the chromosphere, lower corona or TR, and 193 passes some TR emission, as discussed in Section \ref{data}, we cannot rule out that the tail PJs contribute some weak coronal heating. But what we can definitely say from our current observations is that tail PJs display significant signatures in TR emission. However, because of the sparsity of tail PJs, a significant contribution of tail PJs to coronal heating above sunspot penumbrae is very unlikely. High resolution data from future Hi-C flights, Solar-C and DKIST solar observations should be able to address these questions more closely.

\begin{figure*}[tp]
      \centering
      \includegraphics[trim=0cm 3cm 0cm 1cm,clip=true,width=0.98\textwidth]{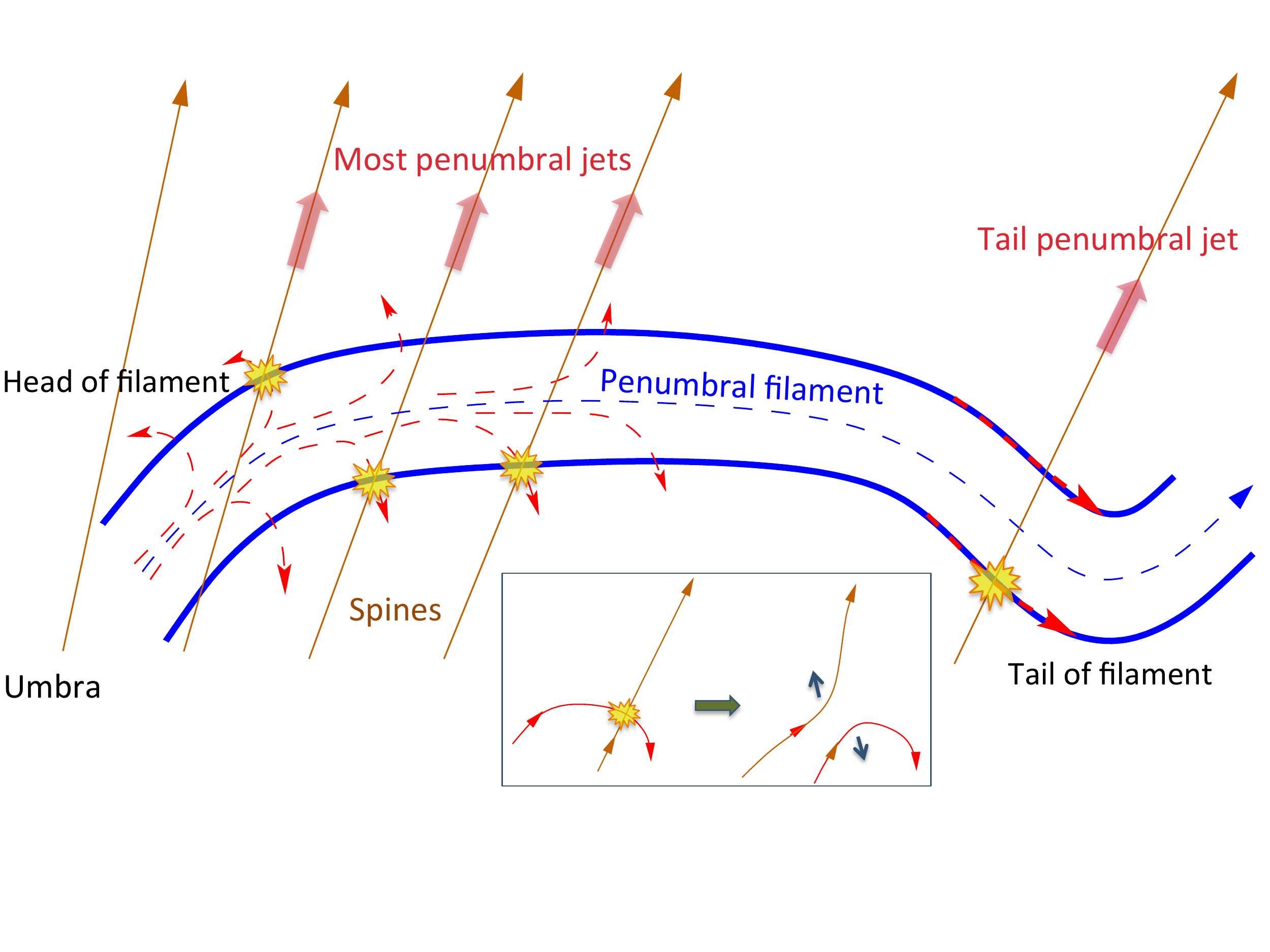}
      \caption{A cartoon diagram (not showing the true width of the filament) depicting the formation mechanism of penumbral jets. For most PJs (the normal PJs), the reconnections, represented by yellow stars, take place at the edges of a penumbral filament, where the field (dashed red lines) is directed at a right angle or obtuse angle to the spine field (dark orange nearly vertical lines). This is a modified picture of that proposed by \cite{kats07}, where reconnection takes place between two components of field inclined to each other at an acute angle. Tail PJs form at tails of filaments, where more field of opposite polarity (to that of the spine field) is present than elsewhere along the filament. All of the larger-than-normal PJs are tail PJs. The proposed magnetic configuration of the reconnection is closely shown inside the box. To keep the picture clearer, the more vertical component of field inside the head of filament, which has nearly the same inclination angle as of the surrounding spine field, is not drawn.} \vspace{0.8cm}
      \label{f1}
\end{figure*}

\subsection{Modified picture of formation of penumbral microjets}\label{sec1}

A sunspot penumbra contains penumbral filaments and spines \citep{tiw15aa}. The recently explored internal structure of sunspot penumbral filaments by \cite{tiw13} suggests the following modification of the formation mechanism of PJs proposed by \cite{kats07}. The penumbral filaments are horizontally elongated magnetized convective cells in which lateral downflows are present. These downflowing lanes also contain opposite polarity field, which is most clearly present near heads of penumbral filaments but also continues to show at sides along the full length of filaments getting strongest at their tails \citep{tiw13}. Although exact location of PJs has not yet been established \citep{jurc08}, PJs possibly occur all along penumbral filaments. On each side of a penumbral filament there is spine field that presses against the filament \citep{tiw13}. The spine field could easily reconnect with the opposite polarity field in the sides of filaments, and produce PJs that travel along the spine field. 

Please note that, with distance from the umbra, the density of spines and field strength in them decrease and their field inclination with respect to vertical increases \citep{tiw15aa}. Thus, the rate of production of PJs and their strength are expected to decrease in the outer penumbra. In the current work, we discovered several locations in penumbra where repeated larger PJs are formed. These locations appear to be the tails of penumbral filaments. We propose that the reconnection between the spine field and tail field can generate these larger PJs. They are large and repetitive because the opposite polarity flux is larger in strength and area there as compared to that on the sides of penumbral filaments.      

In Figure \ref{f1}, we draw a cartoon diagram of the magnetic configuration resulting in the formation of most/normal PJs and tail/larger PJs via magnetic reconnection discussed above. As displayed in the cartoon, the direction of the PJs is still the same as proposed by \cite{kats07}, see also \cite{jurc08}, along the spine field, the more vertical field lines, which changes inclination with radius getting more inclined outward. This picture fits well with the most recent observations of fine-scale structures of sunspot penumbrae. The earlier picture of \cite{kats07} of formation of PJs by reconnection between the two magnetic flux tubes inclined at an acute angle to each other does not fit the recent observations of the convective nature of penumbral filaments. The magnetic field along a filament's axis in the head of the filament and that in the surrounding spines have nearly similar inclination, so reconnection between the spines and opposite polarity field at the edges of penumbral filaments is more likely than it taking place between the field of spines and the same polarity field inside the heads and along the bulk of filaments. In the present picture, at the sides of the penumbral filaments, where the angle between the spine field and the filament field is greatest, most PJs form. Depending on the size of tails of filaments, or a convergence of tails of more than one filament, magnetic reconnection can repetitively produce larger/tail PJs, in the way shown in the cartoon.  
 
The presence of patches of opposite polarity field in the penumbra at the locations of tail PJs is also compatible with the sea-serpent field \citep[e.g.,][]{sain08}. Although the reconnection of the magnetic field in the tails of penumbral filaments with the opposite polarity spine field seems to be the most plausible mechanism of formation of larger PJs, the reconnection of tail fields with the heads of neighboring filaments generating larger PJs cannot be ruled out from our observations. Neither can the emergence of bipolar field in the penumbra leading to such larger PJs via magnetic reconnection with the existing penumbral field be ruled out. It is also unknown whether most PJs form at sites of flux cancellation or emergence; both of these processes are appropriate for producing them. Our picture needs to be verified by higher resolution observations of future observatories, e.g., Solar-C and DKIST. 

Our proposal is supported by the reconnection produced by \cite{maga10} in MHD modeling of PJs, see also modeling by \cite{saka08}. In their model penumbral filaments are assumed to be one twisted flux tube. We believe that this model corresponds to either side of a penumbral filament (Figure \ref{f1}). The reconnection between opposite polarity field at one side of twisted magnetic flux tube can produce jets in a similar way as our proposal for either side of real penumbral filaments. 

It is important to mention here that only one third of filaments selected by \cite{tiw13} displayed opposite polarity field in their sides near heads; the number slighly reduces in the middle (to before the tail) of filaments. The fact that this structure is close to the resolution limit of the telescope, and the opposite polarity fields can only be detected by advanced processing of the SP data, e.g., by spatially coupled inversion code \citep{van12,van13,tiw13}, or by deconvolution methods \citep{ruiz13}, or can only be seen in the specially processed ground based observations \citep{scha13}, the possibility of each of the filaments having opposite polarity field not only near their heads but all along the filament cannot be ruled out. In that case, it is possible that the filaments produce PJs all along their sides by the same process as described above, but in some cases we cannot detect them for them being very narrow and small, below the limit of resolution of the present telescopes. 


\section{CONCLUSIONS}

The normal PJs show at most weak signatures in the transition region (in AIA 1600 \AA\ images) and none in the corona: neither in any of the AIA coronal channels nor in the high-resolution Hi-C images of 193 \AA. Weak signatures of largest normal PJs in 1600 \AA\ suggest that a few largest normal PJs directly power some localized transient TR heating.

A few locations of larger penumbral jets are found in the penumbra, which are locations of mixed polarity magnetic flux near tails of filaments in the penumbra. Larger/tail PJs are brighter (up to 60\% more than that of the background penumbra), wider (up to 600 km) and faster (can be more than 250 \kms) than most/normal PJs, but have lifetimes and apparent lengths similar to or at larger ends of that of most PJs. They flash repeatedly at the same location and clearly display signatures in the transition region (in AIA 1600, 304, 171 and 193 channels). However, no pure coronal signature in AIA 94 \AA\ is detected. None of the penumbral jets seem to be heated enough to display a signature in the hot corona, however, all of the larger ones do appear in transition region emission. These results should be verified with higher resolution data of future Hi-C flights and Solar-C mission.

In aggregate, most PJs and tail PJs apparently do not directly produce appreciable coronal heating (normal PJs because of lack of their coronal signatures, larger PJs because of their sparsity and/or their lack of pure coronal signatures), but conceivably contribute significantly to coronal heating via braiding of the coronal field rooted in and around them, or by production of Alfv\'en waves.

We propose a modified picture of the formation mechanism of PJs considering the recent advancements on the observed magnetic structure of sunspot penumbral filaments. It is more likely that PJs are formed by magnetic reconnection between spines and opposite polarity field at the edges of penumbral filaments along each side of penumbral filaments, as found partly in the MHD simulations of \cite{maga10}, as compared to the earlier proposed reconnection between two inclined components of the same polarity field. The larger/tail PJs appear to form as a result of magnetic reconnection between the opposite polarity field of spines and of filament tails.  


\acknowledgments
We are grateful to the referee for constructive comments, which resulted in major modification and improvement of the paper. Hinode is a Japanese mission developed and launched by ISAS/JAXA, collaborating with NAOJ as a domestic partner, NASA and STFC (UK) as international partners. Scientific operation of the Hinode mission is conducted by the Hinode science team organized at ISAS/JAXA. This team mainly consists of scientists from institutes in the partner countries. Support for the post-launch operation is provided by JAXA and NAOJ (Japan), STFC (U.K.), NASA, ESA, and NSC (Norway). The AIA and HMI data are courtesy of NASA/SDO and the AIA and HMI science teams. MSFC/NASA led the Hi-C mission and partners include the SAO in Cambridge, Mass.; LMSAL in Palo Alto, Calif.; the UCLan in Lancashire, England; and the LPIRAS in Moscow. SKT is supported by appointment to the NASA Postdoctoral Program at the NASA/MSFC, administered by ORAU through a contract with NASA. For this work SEA was supported by the National Science Foundation under Grant No. AGS-1157027. ARW and RLM are supported by funding from the LWS TRT Program of the Heliophysics Division of NASA's SMD.\\


\end{document}